\documentstyle[preprint,prd,aps]{revtex}
\global\arraycolsep=2pt
\makeatletter
\def\fmslash{\@ifnextchar[{\fmsl@sh}{\fmsl@sh[0mu]}}
\def\fmsl@sh[#1]#2{%
  \mathchoice
    {\@fmsl@sh\displaystyle{#1}{#2}}%
    {\@fmsl@sh\textstyle{#1}{#2}}%
    {\@fmsl@sh\scriptstyle{#1}{#2}}%
    {\@fmsl@sh\scriptscriptstyle{#1}{#2}}}
\def\@fmsl@sh#1#2#3{\m@th\ooalign{$\hfil#1\mkern#2/\hfil$\crcr$#1#3$}}
\makeatother

\begin{document}
\draft\pagenumbering{roma}
%%%%%%%%%%%%%%%%%%%%%%%%%%%%%%%%%%%%%%%%%%%%%%%%%%%%%%%%%%%%%%%%%%%%%%%%%%%%%%
\author{Yuan-Ben Dai$^c$,   Ming-Qiu Huang$^{a,b}$}
\address{$^a$ CCAST (World Laboratory) P.O. Box 8730, Beijing, 100080}
\address{$^b$ Institute of High Energy 
Physics, Academia Sinica, P.O.Box 918, Beijing 100039, China}
\address{$^c$ Institute of Theoretical Physics, Academia Sinica, 
P.O.Box 2735, Beijing 100080, China} 
\title{ Semileptonic $B$ decays into excited charmed mesons \\from QCD sum rules  }
\date{May 4, 1998}
\maketitle
\thispagestyle{empty}
\vspace{15mm}
\begin{abstract}
 Exclusive semileptonic $B$ decays into excited charmed mesons  
are studied with QCD sum rules in the leading order of heavy quark effective
theory.  Two universal Isgur-Wise functions $\tau$ and $\zeta$ for semileptonic B decays into four lowest 
lying excited $D$ mesons ($D_1$, $D_2^*$, $D'_0$, and $D'_1$) are
determined. The decay rates and branching ratios for these processes are calculated.
\end{abstract}
\vspace{4mm}
 
\vspace{3.cm}
%\noindent
%April 1998
\newpage
\pagenumbering{arabic}
%%%%%%%%%%%%%%%%%%%%%%%%%%%%%%%%%%%%%%%%%%%%%%%%%%%%%%%%%%%%%%%%%%%%%%%%%%%%%%%%
 
\section{Introduction}
\label{sec1} 

The heavy quark effective theory (HQET) \cite{HQET,neubert1} is a useful tool
to describe the spectroscopy and matrix elements of mesons containing a heavy quark.
In the infinite mass limit, the spin and parity of the heavy quark and that of the 
light degrees of freedom are separately conserved. Coupling the spin of light degrees of 
freedom $j_\ell$ with the spin of heavy quark $s_Q=1/2$ yields a doublet of meson
states with total spin $j=j_\ell\pm 1/2$. The ground state mesons with $Q\,\bar q$ 
flavor quantum numbers contain light degrees of freedom with spin-parity
$j_\ell^{P}=\frac12^-$, giving a doublet  ($0^-$,$1^-$).  For $Q=c$ these mesons 
are the doublet ($D$,$D^*$), while $Q=b$ gives the
doublet ($B$,$B^*$). The excited charmed mesons with $j_\ell^P=1^+/2$ and $3^+/2$
are two spin symmetry doublets ($0^+$,$1^+$) and ($1^+$,$2^+$). These mesons are
doublets ($D'_0$,$D'_1$) and ($D_1$,$D^*_2$). The $D_1$ and $D^*_2$ mesons
have been observed with rather small widths, while $D'_0$ and $D'_1$
have not been observed for the reason of being too broad.
The properties of these lowest lying excited  mesons have attracted attention 
in recent years.  
The mass and decay widths have been studied with potential model
\cite{eichten,koko},  relativistic Bethe-Salpeter equation \cite{dai1} and 
 QCD sum rules \cite{colangelo,huang,huang2}.  

Progress has been achieved recently in the studies of semileptonic $B$ decays
into excited charmed mesons ($D'_0$, $D'_1$, $D_1$ and $D^*_2$).
 Semileptonic $B$ decay into an excited heavy meson has 
been observed in experiments \cite{CLEO,CLEO1,ALEPH,ALEPH1}. With some assumptions, CLEO 
and ALEPH collaborations have reported respectively the branching
ratios ${\mathcal B}(B\to D_1\,e\,\bar\nu_e) = (0.56 \pm 0.13\pm 0.08\pm 0.04)\%$ and 
${\mathcal B}(B\to D_1\,e\,\bar\nu_e) = (0.74 \pm 0.16)\%$, as well as the limits 
${\mathcal B}(B\to D_2^*\,e\,\bar\nu_e)<0.8\%(90\%$ C.L) and  
${\mathcal B}(B\to D_2^*\,e\,\bar\nu_e)<0.2\%$\cite{CLEO1,ALEPH1}.
Theoretically,  HQET provides a systematic method for investigating such processes. 
The semileptonic $B$ decay rate to an excited charmed meson is determined by
the corresponding matrix elements of the weak axial-vector and vector currents.
Heavy quark symmetries
can be used to reduce the form factors  parameterizing the matrix elements. In the
infinite quark mass limit these matrix elements are described respectively
by one universal Isgur-Wise function and vanish at zero recoil \cite{IWsr,Leib}.
Extensive investigation in \cite{Leib} shows that
the leading $1/m_Q$ correction at zero recoil can be calculated in a model
independent way in terms of the masses of charmed meson states.

The universal function embodies details of low energy strong interactions
and cannot be calculated from first principles. It must be calculated
in some nonperturbative approaches.
For that purpose,  there are many  viable approaches, including different quark models 
\cite{godfrey,iw2,cccn,wambach,veseli,oliver,DDG}, relativistic Bethe-Salpeter 
equation \cite{dai2}  and QCD sum rules 
\cite{BBBG,shifman,neubert,c-sum}.
In the present work we study the $B$ semileptonic decays to excited charmed meson states
($D'_0$,$D'_1$) and ($D_1$,$D^*_2$) with QCD sum rule in the leading order
of HQET. In particular we compute the relevant universal
Isgur-Wise functions that describe such decays in the $m_Q\to\infty$ limit.

 The remainder of this paper is organized as follows. In Section 
 \ref{sec2}  we present the formulae of weak current matrix elements and 
 of decay rates. In Section 
 \ref{sec3} we begin with a brief review on the interpolating currents 
 for excited heavy mesons and on relevant sum rules for two-point correlators,
 then the sum 
 rules for  Isgur-Wise functions $\tau$ and $\zeta$ are derived. 
 Section \ref{sec4} is devoted to numerical results and implications.
  Finally in Section \ref{sec4} we draw our conclusions.

%%%%%%%%%%%%%%%%%%%%%%%%%%%%%%%%%%%%%%%%%%%%%%%%%%%%%%%%%%%%%%%%%%%%%%%%%%%%% 
\section{Analytic formulae  for semileptonic decay amplitudes  
$B\to (D_1,D^*_2)\ell\bar\nu$ and $B\to (D'_0,D'_1)\ell\bar\nu$}
\label{sec2}

The theoretical description of semileptonic decays involves the
 matrix elements of vector and axial vector currents
 ($V^\mu=\bar c\,\gamma^\mu\,b$ and $A^\mu=\bar c\,\gamma^\mu\gamma_5\,b$)
between $B$ mesons and excited $D$ mesons. For the processes $B\to D_1\ell\bar\nu$ and
$B\to D_2^*\ell\bar\nu$,
these matrix elements can be parameterized as
\begin{mathletters}\label{matrix1}
\begin{eqnarray}%\label{matrix1}
{\langle D_1(v',\epsilon)|\, V^\mu\, |B(v)\rangle}
  &=& \sqrt{m_{D_1}\,m_B} \;[f_{V_1}\, \epsilon^{*\mu} 
  + (f_{V_2} v^\mu + f_{V_3} v'^\mu)\, \epsilon^*\cdot v ]\,,  \\*
{\langle D_1(v',\epsilon)|\, A^\mu\, |B(v)\rangle  }
  &=& i\, \sqrt{m_{D_1}\,m_B}\;f_A\, \varepsilon^{\mu\alpha\beta\gamma} 
  \epsilon^*_\alpha v_\beta v'_\gamma \,,  \\*
{\langle D^*_2(v',\epsilon)|\, A^\mu\, |B(v)\rangle }
  &=&\sqrt{m_{D_2^*}\,m_B}\; [k_{A_1}\, \epsilon^{*\mu\alpha} v_\alpha 
  + (k_{A_2} v^\mu + k_{A_3} v'^\mu)\,
  \epsilon^*_{\alpha\beta}\, v^\alpha v^\beta ]\,,  \\*
{\langle D^*_2(v',\epsilon)|\, V^\mu\, |B(v)\rangle}
  &=& i\,\sqrt{m_{D_2^*}\,m_B}\;k_V\, \varepsilon^{\mu\alpha\beta\gamma} 
  \epsilon^*_{\alpha\sigma} v^\sigma v_\beta v'_\gamma \,.  \label{mx1-1}
\end{eqnarray}\end{mathletters}
The form factors $f_i$ and $k_i$ are   functions of $y=v\cdot v'$.
In the limit $m_Q\to\infty$ they  can be expressed by  a universal
dimensionless function 
$\tau (y)$~\cite{IWsr,Leib}.
\begin{eqnarray}
   && f_{V_1}(y) ={1\over\sqrt{6}}\,(1-y^2)\,\tau (y) \,,\hspace{1.cm}
     f_{V_2}(y) =  -{3\over\sqrt{6}}\, \tau (y) \,, \nonumber\\
   && f_{V_3}(y) =  {1\over\sqrt{6}} \,(y-2)\, \tau (y) \,,  \hspace{1.13cm}
    f_A(y) = -{1\over\sqrt{6}} \,(1+y)\,\tau (y) \,,
    \nonumber\\
   &&k_{A_1}(y) = -(1+y)\, \tau (y) \,, \hspace{1.53cm} 
    k_{A_2}(y) =  0 \,,
    \nonumber\\
  &&  k_{A_3}(y) = \tau (y) \,, \hspace{3.12cm}  
    k_V(y) = - \tau (w) \,.
\label{relation1}
\end{eqnarray}
Since the polarisation vector of the spin-one state $D_1$ and the 
polarization tensor of the spin-two state $D_2^*$
satisfy $\epsilon^*\cdot v'=0$  and $\epsilon^{*\mu\alpha} v'_\alpha=0$
respectively, only the form factor
$f_{V_1}$   contributes at zero recoil ($v=v'$).

 The form factors that parameterize the
$B\to D'_0\ell\bar\nu$ and $B\to D'_1\ell\bar\nu$ matrix elements of 
the weak currents are defined by
\begin{mathletters}\label{matrix2}
\begin{eqnarray}%\label{matrix2}
\langle D'_0(v')|\, V^\mu\, |B(v)\rangle &=& 0,  \\*
{\langle D'_0(v')|\, A^\mu\, |B(v)\rangle }
  &=& \sqrt{m_{D'_0}\,m_B}\;\big[g_+\, (v^\mu+v'^\mu) + g_-\, (v^\mu-v'^\mu)\big] \,,
    \\*
{\langle D'_1(v',\epsilon)|\, V^\mu\, |B(v)\rangle}
  &=& \sqrt{m_{D'_1}\,m_B}\;\big[g_{V_1}\, \epsilon^{*\, \mu} 
  + (g_{V_2} v^\mu + g_{V_3} v'^\mu)\, (\epsilon^*\cdot v)\big] \,,   \\*
{\langle D'_1(v',\epsilon)|\, A^\mu\, |B(v)\rangle  }
  &=&  \sqrt{m_{D'_1}\,m_B}\;i\,g_A\, \varepsilon^{\mu\alpha\beta\gamma}\, 
  \epsilon^*_\alpha v_\beta\, v'_\gamma \,,  
\end{eqnarray}\end{mathletters}
where $g_i$ are functions of $y$.  At zero recoil the matrix elements are
determined by $g_+(1)$ and $g_{V_1}(1)$.  In the infinite mass limit form factors
 $g_i$  can be written 
 in terms  of a single function $\zeta(y)$ \cite{IWsr,Leib},
  \begin{eqnarray}                                                                       
   && g_{+}(y) = 0 \,,\hspace{1.2cm}
     g_{-}(y) = \zeta (y) \,, \nonumber\\
    &&g_{A}(y) =  \zeta(y) \,, \hspace{0.79cm} 
    g_{v_1}(y) =  (y-1)\,\zeta(y)\,,
    \nonumber\\
  &&  g_{v_2}(y) =  0 \,, \hspace{1.2cm}  
    g_{V_3}(y) = - \zeta(w) \,.
\label{relation2}
\end{eqnarray}
Note that  the notations of Ref.~\cite{Leib} have been used here. $\tau$ 
is $\sqrt3$ times the function $\tau_{3/2}$ of Ref.~\cite{IWsr}, while 
$\zeta$ is two times the function of $\tau_{1/2}$.
  All of these matrix elements of the weak currents vanish at zero recoil  $y=1$,
since the $B$ meson and the $(D_1,D_2^*)$ or $(D'_0,D'_1)$ mesons are in
different heavy quark spin symmetry multiplets, and the current at zero recoil
is related to the conserved charges of heavy quark spin-flavor symmetry. 

The differential   decay rates are given by (taking the mass of the final 
lepton to zero)
\begin{eqnarray}
   \frac{{\mathrm d}\Gamma_{D_1}}{{\mathrm d}y}
   &=& \frac{G_F^2\,|V_{cb}|^2 m_B^5}{72\pi^3}\,r_1^3 \,(y+1){(y^2-1)}^{3/2}
     \left[ (y-1)(1+r_1)^2   +y(1-2r_1y +r_1^2)\right] \vert
\tau (y)\vert^2  \,,\label{rate1}\\
   \frac{{\mathrm d}\Gamma_{D^*_2}}
    {{\mathrm d}y}
   &=& \frac{G_F^2\,|V_{cb}|^2 m_B^5}{ 72\pi^3}\,r_2^3\,(y+1)
    (y^2-1)^{3/2}   \left[  (y+1)  (1-r_2)^2 +3 y (1-2r_2y +r_2^2) 
     \right]  \vert \tau (y)\vert^2  \,,
\label{rate2}\\
\frac{{\mathrm d}\Gamma_{D'_0}}{{\mathrm d}y}
   &=& \frac{G_F^2\,|V_{cb}|^2 m_B^5}{48\pi^3}\,r_0^{'3 }{(y^2-1)}^{3/2}\;
        (1-r'_0)^2 \vert \zeta(y)\vert^2  \label{rate3} \;,\\
\frac{{\mathrm d}\Gamma_{D'_1}}
    {{\mathrm d}y}
   &=& \frac{G_F^2\,|V_{cb}|^2 m_B^5}{48\pi^3}\,r_1^{'3}\,
   (y-1) \sqrt{y^2-1}  \left[  (y-1) (1+r'_1)^2+ 4y(1-2r'_1 y+r_1^{'2})
    \right]\vert \zeta(y)\vert^2 \,,
\label{rate4}
\end{eqnarray}
where   $r_1=m_{D_1}/m_B$, $r_2=m_{D_2^*}/m_B$, $r'_0=m_{D'_0}/m_B$ and 
$r'_1=m_{D'_1}/m_B$.

%%%%%%%%%%%%%%%%%%%%%%%%%%%%%%%%%%%%%%%%%%%%%%%%%%%%%%%%%%%%%%%%%%%%%%%%%%%%%%%%
\section{Sum rules for Isgur-Wise functions $\tau$ and $\zeta$}
\label{sec3}
%%%%%%%%%%%%%%%%%%%%%%%%%%%%%%%%%%%%%%%%%%%%%%%%%%%%%%%%%%%%%%%%%%%%%%%%%%%%%%%
\subsection{Interpolating currents for heavy mesons of arbitrary spin and parity
and two-point correlation function}

A basic element in the application of QCD sum rules to excited heavy mesons is 
to choose a set of appropriate interpolating currents  in terms of quark fields 
each of which creates
(annihilate) a definite excited state of the heavy mesons.  The proper interpolating 
current $J_{j,P,j_{\ell}}^{\alpha_1\cdots\alpha_j}$
for the state with the quantum number $j$, $P$, $j_{\ell}$ in HQET was
given in \cite{huang}. 
These currents have nice properties. 
They were proved to satisfy the following conditions 
\begin{eqnarray}
\label{decay}
\langle 0|J_{j,P,j_{\ell}}^{\alpha_1\cdots\alpha_j}(0)|j',P',j_{\ell}^{'}\rangle&=&i\,
f_{Pj_l}\delta_{jj'}
\delta_{PP'}\delta_{j_{\ell}j_{\ell}^{'}}\eta^{\alpha_1\cdots\alpha_j}\;,\\
\label{corr}
i\:\langle 0|T\left (J_{j,P,j_{\ell}}^{\alpha_1\cdots\alpha_j}(x)J_{j',P',j_{\ell}'}^{\dag
\beta_1\cdots\beta_{j'}}(0)\right )|0\rangle&=&\delta_{jj'}\delta_{PP'}\delta_{j_{\ell}j_{\ell}'}
(-1)^j\:{\cal S}\:g_t^{\alpha_1\beta_1}\cdots g_t^{\alpha_j\beta_j}\nonumber\\[2mm]&&\times\:
\int \,dt\delta(x-vt)\:\Pi_{P,j_{\ell}}(x)
\end{eqnarray}
in the $m_Q\to\infty$ limit. Where $\eta^{\alpha_1\cdots\alpha_j}$ is the 
polarization tensor for the spin $j$ state,  $v$ is the velocity of the heavy
quark, $g^{\alpha\beta}_t=g^{\alpha\beta}-v^\alpha v^\beta$ is the transverse 
metric tensor, ${\cal S}$ denotes symmetrizing the indices and
subtracting the trace terms separately in the sets $(\alpha_1\cdots\alpha_j)$
and $(\beta_1\cdots\beta_{j})$, $f_{P,j_{\ell}}$ and $\Pi_{P,j_{\ell}}$ are
a constant and a function of $x$ respectively which depend only on $P$ and $%
j_{\ell}$. Because of equations (\ref{decay}) and (\ref{corr}), the sum rules
in HQET for decay amplitudes derived from a correlator containing such currents
receive contribution only from one of the two states with the same spin-parity
$(j,P)$ in the $m_Q\to\infty$. Starting from the
calculations in the leading order, the decay amplitudes for finite $m_Q$ can
be calculated unambiguously order by order in the $1/m_Q$ expansion in HQET.

In the following we focus our attention on the semileptonic decays,
$B\to D_1$, $D^*_2$  and $B\to  D'_0$, $D'_1$. The relevant doublets to be
considered are ground states and lowest lying positive parity states,
namely, doublets ($0^-$,$1^-$), ($0^+$,$1^+$) and ($1^+$,$2^+$). The currents 
for creating $0^-$ and $1^-$ are usual pseudoscalar and vector currents
\begin{eqnarray}
\label{p-vector}
J^{\dag\alpha}_{0,-,{1\over 2}}=\sqrt{\frac{1}{2}}\:
\bar h_v\gamma_5q\;, \hspace{1.5cm} J^{\dag\alpha}_{1,-,{1\over 2}}=
\sqrt{\frac{1}{2}}\:\bar h_v\gamma_t^{\alpha} q\;. 
\end{eqnarray}
As pointed out in \cite{huang},   there are 
two possible choices for currents creating $0^+$ and $1^+$ of the doublet
 $(0^+,1^+)$, either
\begin{eqnarray}
\label{curr1}
J^{\dag}_{0,+,2}&=&\frac{1}{\sqrt{2}}\:\bar h_vq\;,\\
\label{curr2}
J^{\dag\alpha}_{1,+,2}&=&\frac{1}{\sqrt{2}}\:\bar h_v\gamma^5\gamma^{\alpha}_tq\;,
\end{eqnarray}
or
\begin{eqnarray}
\label{curr3}
J^{'\dag}_{0,+,2}&=&\frac{1}{\sqrt{2}}\:\bar h_v(-i)\fmslash{\cal D}_tq\;,\\
\label{curr4}
J^{'\dag}_{1,+,2}&=&\frac{1}{\sqrt{2}}\:\bar h_v\gamma^5\gamma^{\alpha}_t(-i)
\fmslash{\cal D}_tq\;,
\end{eqnarray}
where ${\cal D}$ is the covariant derivative. Similarly, there are two possible choices for the currents creating $1^+$ and $2^+$ 
of 
the doublet $(1^+,2^+)$. One is  
\begin{eqnarray}
\label{curr5}
J^{\dag\alpha}_{1,+,1}&=&\sqrt{\frac{3}{4}}\:\bar h_v\gamma^5(-i)\left(
{\cal D}_t^{\alpha}-\frac{1}{3}\gamma_t^{\alpha}\fmslash{\cal D}_t\right)q\;,\\
\label{curr6}
J^{\dag\alpha_1,\alpha_2}_{2,+,1}&=&\sqrt{\frac{1}{2}}\:\bar h_v
\frac{(-i)}{2}\left(\gamma_t^{\alpha_1}{\cal D}_t^{\alpha_2}+
\gamma_t^{\alpha_2}{\cal D}_t^{\alpha_1}-\frac{2}{3}\;g_t^{\alpha_1\alpha_2}
\fmslash{\cal D}_t\right)q\;.
\end{eqnarray}
Another choice is obtained by adding a factor $-i\fmslash{\cal D}_t$ to (\ref{curr5})
 and (\ref{curr6}). Note
that, without the last term in the bracket in (\ref{curr5}) the current
would couple also to the $1^+$ state in the doublet $(0^+,1^+)$ even in
the limit of infinite $m_Q$.

Usually, the currents with the least number of derivatives are used in
the QCD sum rule approach. However,   there is certain motivation for using the
currents (\ref{curr3}), (\ref{curr4}) for the doublet $(0^+,1^+)$ for the reason that in the
non-relativistic quark model,  the doublets $(0^+,1^+)$  
are orbital p-wave states which correspond to one derivative in the space
wave functions.   Therefore, we shall consider both the currents
(\ref{curr1}), (\ref{curr2}) and (\ref{curr3}), (\ref{curr4}) for the
doublet $(0^+,1^+)$.

For the ground state heavy mesons, the  sum rule for the correlator of two heavy-light
currents  is well-known. It is: \cite{neubert1,neubert}
\begin{eqnarray}
\label{form}
   f_{-,{1\over 2}}^2\,e^{-2\bar\Lambda_{-,{1\over 2}}/T} &=&
   \frac{3}{16\pi^2}\int_0^{\omega_{c0}}\omega^2e^{-\omega/{T}}
   d\omega - \frac12\langle\bar q q\rangle\left(1-
   \frac{m_0^2}{4 T^2}\right) \,.\label{2-point1}
\end{eqnarray}   
For the doublet $(0^+,1^+)$, when the currents ${J'}_{0,+,2}$, ${J'}_{1,+,2}$ 
in (\ref{curr3}), (\ref{curr4}) are
used the sum rule (same for the two states) after the Borel
transformation is found to be \cite{huang}
\begin{eqnarray}
\label{form1}
&&f_{+,{1\over 2}}^{'2}e^{-2\bar\Lambda'_{+,{1\over 2}}/{T}}=
\frac{3}{2^6\pi^2}\int_0^{\omega'_{c1}}\omega^4e^{-\omega/{T}}d\omega-
\frac{1}{2^4}\,m_0^2\,\langle\bar qq\rangle\;.
\end{eqnarray}
The corresponding formula when the current $J_{0,+,2}$ and $J_{1,+,2}$ 
in (\ref{curr1}) and (\ref{curr2}) are used instead of
${J'}_{0,+,2}$ and ${J'}_{1,+,2}$ is the following \cite{huang}
\begin{eqnarray}
\label{form2}
&&f_{+,{1\over 2}}^{2}e^{-2\bar\Lambda_{+,{1\over 2}}/{T}}=
\frac{3}{16\pi^2}\int_0^{\omega_{c1}}\omega^2e^{-\omega/{T}}d\omega+
\frac{1}{2}\,\langle\bar qq\rangle-{1\over 8T^2}\,m_0^2
\,\langle\bar qq\rangle\;.
\end{eqnarray}
When the currents (\ref{curr5}) and (\ref{curr6}) are used the sum rule for
the $(1^+,2^+)$  doublet is found to be \cite{huang}
\begin{eqnarray}
\label{form3}
&&f_{+,{3\over 2}}^2e^{-2\bar\Lambda_{+,{3\over 2}}/{T}}={1\over 2^6\pi^2}
\int_0^{\omega_{c2}}\omega^4e^{-\omega/{T}}d\omega
-\frac{1}{12}\:m_0^2\:\langle\bar qq\rangle-{1\over 2^5}\langle{\alpha_s\over\pi}G^2\rangle T\;.
\end{eqnarray}
Here $m_0^2\,\langle\bar qq\rangle=\langle\bar qg\sigma_{\mu\nu}G^{\mu\nu}q\rangle$.
  Above results will be used in the following sections.

%%%%%%%%%%%%%%%%%%%%%%%%%%%%%%%%%%%%%%%%%%%%%%%%%%%%%%%%%%%%%%%%%%%%%%%%%%%%%%%%%%%%%%%
\subsection{sum rules for $\tau$ and $\zeta$}
For the ampltudes of the semileptonic decays to excited states in the
infinite mass limit, the only unknown
quantities in the expressions (5)-(8) are the universal  functions 
$\tau(y)$ and $\zeta(y)$.  Let us first consider the Isgur-Wise function $\tau$.
In order to calculate this form factor using QCD sum rules, one studies
the analytic properties of the three-point correlators
\begin{mathletters}\label{3-point1}
\begin{eqnarray}
 i^2\int\, d^4xd^4z\,e^{i(k'\cdot x-k\cdot z)}\;\langle 0|T\left(
 J^{\nu}_{1,+,\frac{3}{2}}(x)\;{\cal J}^{\mu(v,v')}_{V,A}(0)\;   
 J^{\dagger}_{0,-,\frac{1}{2}}(z)\right)|0\rangle&=&\Xi(\omega,\omega',y)\;{\cal L}
 ^{\mu\nu}_{V,A}\;, \\
i^2\int\, d^4xd^4z\,e^{i(k'\cdot x-k\cdot z)}\;\langle 0|T\left(
 J^{\alpha\beta}_{2,+,\frac{3}{2}}(x)\;{\cal J}^{\mu(v,v')}_{V,A}(0)\;   
 J^{\dagger}_{0,-,\frac{1}{2}}(z)\right)|0\rangle&=&\Xi(\omega,\omega',y)\;{\cal L}
 ^{\mu\alpha\beta}_{V,A}\;,
\end{eqnarray}
\end{mathletters}
where ${\cal J}^{\mu(v,v')}_{V}=\bar h(v')\gamma^\mu\,h(v)$, 
${\cal J}^{\mu(v,v')}_{A}=\bar h(v')\gamma^\mu\gamma_5\,h(v)$. 
The variables $k$, $k'$ denote residual ``off-shell" momenta which are related to the
momenta $P$ of the heavy quark in the initial 
state and $P'$ in the final state by $k=P-m_Qv$, $k'=P'-m_{Q'}v'$ 
respectively. For heavy quarks in bound states they
 are typically of order $\Lambda_{QCD}$ and remain finite in the
heavy quark limit.  ${\cal L}_{V,A}$ are  Lorentz structures  associated with
the vector and axial vector 
currents(see Appendix).

The coefficient $\Xi(\omega,\omega',y)$ in (\ref{3-point1}) is an analytic 
function in the ``off-shell energies" $\omega=2v\cdot k$ and $\omega'=2v'\cdot
k'$
with discontinuities for positive values of these variables. It furthermore
depends on the velocity transfer $y=v\cdot v'$, which is fixed at its physical region for
the process under consideration. By
saturating
(\ref{3-point1}) with physical intermediate states in HQET, one finds the
hadronic representation of the correlator as following
 \begin{eqnarray}
\label{pole}
\Xi_{hadro}(\omega,\omega',y)={f_{-,{1\over 2}}f_{+,{3\over 2}}\tau(y)
\over (2\bar\Lambda_{-,{1\over 2}}-\omega- i\epsilon
)(2\bar\Lambda_{+,{3\over 2}}-\omega'- i\epsilon)}+\mbox{higher resonances} \;,
\end{eqnarray}
where $f_{P,j_\ell}$ are constants defined in (\ref{decay}),
$\bar\Lambda_{P,j_\ell}=m_{P,j_\ell}-m_Q$.
 As the result of
equation (\ref{decay}), only one state with $j^P=1^+$ or $j^P=2^+$
 contributes to (\ref{pole}), the other resonance with the same quantum
 number $j^P$ and different $j_l$ does not contribute.
  This would not be true for $j^P=1^+$ if the last term in (\ref
{curr5}) is absent.

Using the expression for the heavy quark propagator $\int dt \delta(x-vt)(1+\fmslash v)/2$ in
HQET the correlators in (\ref{3-point1}) have the following form
\begin{eqnarray}
i^2 \int\;dt_1 dt_2 \;e^{i (\omega t_1+\omega' t_2)/2}\;\mbox{{\bf Tr}}\left\{\frac{1+\fmslash
v}{2}\gamma_{V,A} \frac{1+\fmslash v'}{2}\langle 0|\Gamma' q(v't_2) \bar q(-vt_1)\Gamma
|0\rangle\right\} \;,\nonumber
\end{eqnarray}
where $\Gamma$ and $\Gamma'$ are functions of Dirac matrices and covariant derivatives
appearing in the definations of the currents. For $\omega$ and $\omega'$ in the
deep Euclidian region the integral is dominated by small values of $t_1$ and
$t_2$. As usually done in QCD sum rule approach, $\omega$ and $\omega'$
are analytically continued to the deep Euclidian region. Therefore OPE in the
short distance can be applied to the above expression for any value of $y$ not much
larger than $1$. This is because, for small $t_1$, $t_2$ and $y$ of the order ${\cal O}(1)$, 
 all components of  $vt_1$ and $v't_2$ can be made small by choosing the rest frame of $B$.
In our case the maxima value of $y$ is less than $1.5$.  Therefore
$\Xi(\omega,\omega',y)$ in (\ref{3-point1}) can be approximated by a perturbative
calculation supplemented
by nonperturbative power corrections proportional to the vacuum condensates which
are treated as  phenomenological parameters.
The perturbative contribution can be represented by a double dispersion integral
in $\omega$ and $\omega'$ plus possible subtraction terms. 
We find that the theoretical expression for the
correlator has the form:
\begin{eqnarray}
\label{theo}
\Xi_{theo}(\omega,\omega',y)&\simeq&\int\; d\nu d\nu'\frac{\rho^{pert}(\nu,\nu',y)}
{(\nu-\omega-i\epsilon)(\nu'-\omega'-i\epsilon)}+\mbox{subtractions}\nonumber\\
&& \quad\mbox{}+
\Xi^{cond}(\omega,\omega',y)\;.
\end{eqnarray}
  
Following the standard QCD sum rule procedure the calculations of
$\Xi(\omega,\omega',y)$ are straitforward. Confining us to the leading order of perturbation
and the operators with dimension $D\leq 5$ in OPE, the relevant Feynman
diagrams are shown in Fig 1.  

 The perturbative part of the spectral density is
\begin{eqnarray}
\label{spectral1}
 \rho_{pert}(\tilde\omega,\tilde\omega',y)&=&\frac{3}{2^7\pi^2}
 \frac{1}{(1+y)(y^2-1)^{3/2}}
 \big(-3\tilde\omega^2+(\tilde\omega^{'2}+2\tilde\omega\tilde\omega')
 (2y-1)\big)\;\nonumber\\
&& \quad\mbox{}\times\Theta(\tilde\omega)\,\Theta(\tilde\omega')\,
\Theta(2y\tilde\omega\tilde\omega'-\tilde\omega^2-\tilde\omega^{'2})\;.
\end{eqnarray}

The QCD sum rule is obtained by equating the phenomenological and
theoretical expressions for $\Xi$. In doing this the quark-hadron duality 
needs to be assumed to model the contributions of
higher resonance part of Eq. (\ref{pole}). Generally speaking, 
the duality is to simulate
 the resonance contribution by the perturbative 
part above some threshold energys. In the QCD sum rule analysis for  $B$
semileptonic decays into ground state $D$ mesons, it is argued by
Neubert and Shifman
in \cite{neubert1,shifman,neubert} that the perturbative and the hadronic spectral
densities can not be locally dual to each other, the necessary way to restore
duality  is to integrate the spectral densities over the ``off-diagonal''
variable $\tilde\omega_-=(\tilde\omega-\tilde\omega')/2$, keeping the  ``diagonal'' variable
$\tilde\omega_+=(\tilde\omega+\tilde\omega')/2$ fixed. It is in $\tilde\omega_+$ 
that the quark-hadron duality is assumed for the integrated spectral densities. 
We shall use the same prescription in the case 
of $B$ semileptonic decays into excited state $D$ mesons.

The $\Theta$ functions in (\ref{spectral1}) imply that in terms of $\tilde\omega_+$
and $\tilde\omega_-$ the double
discontinuities of the corrrelator are confined to the region
 $-\sqrt{y^2-1}/(1+y)\;\tilde\omega_+\leq\tilde\omega_-\leq\sqrt{y^2-1}/(1+y)\;\tilde\omega_+$
and $\tilde\omega_+\geq 0$. According to our prescription an isosceles triangle
with the base $\tilde\omega_+ = \tilde\omega_c$ is retained in the integation
domain of the perturbative term in the sum rule.

In view of the asymmetry of the problem at hand with respect to the
initial and final states one may attempt to use an asymmetric triangle
in the perturbative integral. However, in that case the factor
$(y^2-1)^{3/2}$ in the denominator of (\ref {spectral1}) is not canceled
after the integration so that the Isgur-Wise function or it's
derivative will be divergent at $y=1$. Similar situation occurs for the
sum rule of the Isgur-Wise function for tansition between ground states
if a different domain is taken in the perturbative integal \cite{neubert}.

In order to supress the contributions of higher resonance states
a double Borel transformation in $\omega$ and $\omega'$ is performed to both sides
of the sum rule, which introduces two Borel parameters $T_1$ and $T_2$.  
For simplicity we shall take the two Borel parameters equal:  
$T_1 = T_2 =2T$. In the following section we shall estimate 
the changes in the sum rules in the case of $T_1 \neq T_2$.

After adding the non-perturbative part and making the double Borel
transformation one obtains the sum rule for $\tau$ as follows 
\begin{eqnarray}
\label{sum-rule1}
\tau(y)\,f_{-,\frac{1}{2}}\,f_{+,\frac{3}{2}}\;e^{-(\bar\Lambda_{-,\frac{1}{2}}
+\bar\Lambda_{+,\frac{3}{2}})/T}&=&
\frac{1}{2\pi^2}\frac{1}{(y+1)^3}\;\int_0^{\omega_c}d{\omega_+}\,
\omega_+^3\,e^{-\omega_+/T}-\frac{1}{12}m^2_0{\langle\bar qq\rangle\over T}
\nonumber\\
&& \quad\mbox{}-\frac{1}{3\times 2^5}  
\langle \frac{\alpha_s}{\pi}GG\rangle\frac{y+5}{(y+1)^2}\;.
\end{eqnarray}

 We now turn to the study of $\zeta(y)$. To obtain QCD sum rules for  this 
Isgur-Wise function  one starts from three-point correlators
\begin{mathletters}\label{3-point2}
\begin{eqnarray}
 i^2\int\, d^4xd^4z\,e^{i(k'\cdot x-k\cdot z)}\;\langle 0|T\left(
 J_{0,+,\frac{1}{2}}(x)\;{\cal J}^{\mu(v,v')}_{A}(0)\;   
 J^{\dagger}_{0,-,\frac{1}{2}}(z)\right)|0\rangle&=&\Xi(\omega,\omega',y)\;
 {\cal D}^{\mu}_{A}\;, \label{point2-1}\\
i^2\int\, d^4xd^4z\,e^{i(k'\cdot x-k\cdot z)}\;\langle 0|T\left(
 J^{\nu}_{1,+,\frac{1}{2}}(x)\;{\cal J}^{\mu(v,v')}_{V,A}(0)\;   
 J^{\dagger}_{0,-,\frac{1}{2}}(z)\right)|0\rangle&=&\Xi(\omega,\omega',y)\;
 {\cal D}^{\mu \nu}_{V,A}\;,
 \end{eqnarray}
\end{mathletters}
the inserting of vector current in (\ref{point2-1}) vanishes.  

  As mentioned above 
there are two possible choices for the currents creating $0^+$ and $1^+$
of the doublet $(0^+,1^+)$. 
 Let us first consider the sum rule by using  the interpolating currents (\ref{curr3}) and   (\ref{curr4}) 
with  derivative. The perturbative part of the spectra density is found to be
\begin{eqnarray}
\label{spectral2}
 \rho'_{pert}(\tilde\omega,\tilde\omega',y)&=&\frac{3}{2^6\pi^2}
 \frac{y+1}{(y^2-1)^{3/2}}
 \big( \tilde\omega^{'2}- \tilde\omega\tilde\omega'\big)
 \Theta(\tilde\omega)\,\Theta(\tilde\omega')\,
\Theta(2y\tilde\omega\tilde\omega'-\tilde\omega^2-\tilde\omega^{'2})
  \;,\nonumber\\.
\end{eqnarray}
With the same procedure the resulting sum rule for $\zeta$ takes the form
\begin{eqnarray}
\label{zeta1}
\zeta(y)\;f_{-,\frac{1}{2}}\,f'_{+,\frac{1}{2}}\;e^{-(\bar\Lambda_{-,\frac{1}{2}}
+\bar\Lambda'_{+,\frac{1}{2}})/T}&=&
\frac{1}{8\pi^2}\frac{1}{(y+1)^2}\;\int_0^{\omega_c}d{\omega_+}\,
\omega_+^3\,e^{-\omega_+/T}-\frac{1}{12}m^2_0{\langle\bar qq\rangle}{1+y\over T}
\nonumber\\&& \quad\mbox{}
-\frac{1}{3\times 2^6}\langle\frac{\alpha_s}{\pi} GG\rangle\frac{7y+1}{y+1} \;.
\end{eqnarray}

When the currents (\ref{curr1}) and (\ref{curr2}) without derivative are used,
there is no diagram  shown  in Fig (c).  The evaluation of the perturbative
graph gives the   spectral density 
\begin{eqnarray}
\label{spectral20}
 \rho_{pert}(\tilde\omega,\tilde\omega',y)=\frac{3}{2^5\pi^2}
 \frac{1+y}{(y^2-1)^{3/2}}
  (\tilde\omega -\tilde\omega^{'}) \;\Theta(\tilde\omega)\,\Theta(\tilde\omega')\,
\Theta(2y\tilde\omega\tilde\omega'-\tilde\omega^2-\tilde\omega^{'2})\;.
 \;. 
\end{eqnarray}

After rewriting the spectral function in terms of $\tilde\omega_{\pm}$,
 performing the double Borel transformation and
the integral over $\tilde\omega_-$ in the confined region, we find that
the perturbative contribution vanishes for $T_1=T_2$. 
Therefore, the sum rule in this case is not a good one. We shall not use
it in the numerical analysis in the following.  

We end this subsection by noting that the QCD $O(\alpha_s)$ corrections have not been included in the 
sum rule calculations.  
 However,  the Isgur-Wise function obtained from the QCD sum rule 
actually is a ratio of the three-point correlator to the two-point 
correlator results.  While both of these correlators subject to large 
perturbative QCD corrections, it is expected that their ratio is not much by these 
corrections significantly because of cancelation.   
 This has been  proved to be true
in the analysis for $B$ semileptonic decay to ground state 
heavy mesons \cite{neubert}.
 
%%%%%%%%%%%%%%%%%%%%%%%%%%%%%%%%%%%%%%%%%%%%%%%%%%%%%%%%%%%%%%%%%%%%%%%%%%%%%%%
\section{Numerical results and implications }
\label{sec4}

 We now turn to the numerical evaluation of these sum rules and implications.
For the QCD parameters entering the theoretical expressions, we take
 the standard values 
 \begin{eqnarray}
   \langle\bar q q\rangle &=& -(0.23)^3~\mbox{GeV}^3
    \,, \nonumber\\
   \langle\alpha_s GG\rangle &=& (0.04)~\mbox{GeV}^4 \,,
    \nonumber\\
   m_0^2 &=& (0.8)~\mbox{GeV}^2 \,,
\label{cond}
\end{eqnarray}
In the numerical calculations we use the  physical masses for
 $B$, $D_1$ and $D^*_2$ \cite{PDG}. Whereas for $D'_0$ and $D'_1$ we take the theoretical
mass values of the  doublet in the leading order obtained in \cite{huang}.
Therefore we use
\begin{eqnarray}
   m_B=5.279,\quad  m_{D_1}=2.422,\quad m_{D_2^\ast}=2.459,
   \quad m_{D^{\ast}_0}=m_{D'_1}=2.4, 
\label{masses}  
 \end{eqnarray}
as well as $V_{cb}=0.04$. 

 In order to obtain information for $\tau(y)$ and $\zeta(y)$ from the sum rules
 which is independent of specific input values of $f$'s and $\bar\Lambda$'s,
we adopt the strategy to evaluate  the sum rule by  eliminating
the explicit dependence on the parameters $f$'s and $\bar\Lambda$'s by
using the  sum rules for the correlator of two heavy-light
currents.
Dividing the three point sum rules in (\ref{sum-rule1}) and (\ref{zeta1})
by the square roots of relevant  two point sum rules in
(\ref{2-point1}), (\ref{form1}) and (\ref{form3}), we obtain expressions for the
  $\tau$ and $\zeta$ as functions of the Borel parameter $T$ and the continuum
thresholds. This procedure may reduce the systematic
uncertainties in the calculation.

 Let us   evaluate numerically the sum rules  for $\tau(y)$ and $\zeta(y)$.
%%dividing the three-point sum rule (\ref{sum-rule1}) with two-point sum rules (\ref{2-sum1})
%%and (\ref{2-sum2})   to find stability value of $\tau$ with respect to the Borel parameter $T$. 
Imposing usual criterium for the upper and lower bounds of the Borel parameter,
we found they have common sum rule ``window'': $0.7<T<1.1$, which overlaps with those
of two-point sum rules \cite{neubert,huang}. Notice that in the case of 
 transiton between ground states the normlization of the Isgur-Wise function
 at zero recoil leads to the requirement that the Borel parameter 
%%$T=T_1/2=T_2/2$ in the sum rule the three point function is equal to the Borel parameter 
%%in the sum rule for the two point function.
 $T_1=T_2\equiv 2T$. That is, the Borel parameter in the sum rule for three-point 
 correlators is twice the Borel parameter in the sum rule for the two-point correlators.
  In Fig. 2, we show the range of predictions for $\tau(y)$ and $\zeta(y)$
obtained by varying the continuum thresholds $\omega_c$, $\omega_{c0}$, $\omega'_{c1}$ 
and $\omega_{c2}$, and fixing the Borel parameter at $T=0.9$.
In the evaluation we have taken $2.0<\omega_c<2.5$, $1.9<\omega_{c0}<2.4$, 
$2.3<\omega'_{c1}<2.8$
and $2.5<\omega_{c2}<3.0$. Where $\omega_{c0}$, $\omega'_{c1}$ and $\omega_{c2}$
are relevant two-point sum rule continuum thresholds defined in (\ref{form}),
(\ref{form1}) and (\ref{form3}),  their ranges
 are determined  by two-point sum rule analysises
\cite{neubert,huang}. The results are stable against reasonable variations of these
parameters.

The resulting curves for $\tau(y)$ and $\zeta(y)$ may be well parameterized by
 the linear approximations
\begin{eqnarray}
\tau(y)&=&\tau(1)\;(1-\rho_{\tau}^2(y-1))\;, \hspace{0.3cm}
\tau(1)=0.74\pm 0.15\;, \hspace{0.3cm} \rho_{\tau}^2=0.90\pm 0.05 \;,\\
\zeta(y)&=&\zeta(1)\;(1-\rho_{\zeta}^2(y-1))\;, \hspace{0.3cm}
\zeta(1)=0.26\pm 0.08\;, \hspace{0.3cm} \rho_{\zeta}^2=0.50\pm 0.05\;.
\end{eqnarray}
The errors reflect the uncertainty due to $\omega$'s and $T$.

 We have estimated the variations of Isgur-Wise functions
at zero recoil in the case of $T_1\neq T_2$.
After Borel transformation the weight function in the dispersion integrals
is $exp\{-(1/T_1+1/T_2)\tilde\omega_+-(1/T_1-1/T_2)\tilde\omega_-\}$.
We found that for $T_2 = 3T_1/2$ the value of $\tau(1)$ increases about $10\%$,
while the value of
$\zeta(1)$  increases about $30\%$. The variations are slower for $T_1$ larger
than $T_2$. This gives us the uncertainties of the results within reasonable
region of $T_1/T_2$.

The numerical values of $\tau$ and $\zeta$ at zero recoil are compared
with other approaches in Table \ref{tab:comp}.
For $\tau$ we find a broad 
agreement with some of the constituent quark model results, whereas for 
$\zeta$ we only agree with \cite{DDG}.
 
 Using the forms of linear approximations for  $\tau(y)$ and $\zeta(y)$ 
 we can compute the total semileptonic rates and branching rations.  
 The maximal values of $y$ in the four cases are 
 $y^{D_1}_{max}=(1+r_1^2)/2r_1\approx 1.32$,  
$y^{D^*_2}_{max}=(1+r_2^2)/2 r_2\approx 1.31$ and
$y^{D'_0}_{max}=y^{D'_1}_{max}=(1+r_0^{'2})/2r'_0\approx 1.33$, respectively.
The results turn out to be
\begin{eqnarray}
\Gamma_{D_1}&=&1.36\times 10^{-15}~~\mbox{GeV}\;,\nonumber\\
\Gamma_{D_2}&=&2.11\times 10^{-15}~~\mbox{GeV}\;,\nonumber\\
 \Gamma_{D'_0}&=&7.37\times 10^{-17}~~\mbox{GeV}\;,\nonumber\\
\Gamma_{D'_1}&=&9.98\times 10^{-17}~~\mbox{GeV}\;.
\end{eqnarray}
  We  found that the semileptonic decay into $j_\ell=1^+/2$ is strongly suppressed.
  The decay widths to  it are about
one order of magnitude smaller than ones into $j_\ell=3^+/2$.  
This predicted suppression needs to be checked in the future experiment.

 In Table \ref{tab:branch} we present our results for the branching ratios of
B semileptonic decays to    lowest lying positive parity charmed mesons. We have
taken $\tau_B=1.62$ ps.
The ratio of the two semileptonic rates for $B$ decays into $D_1$ and $D_2^*$
mesons  is
\begin{equation}
R \equiv {{\cal B}(B\to D_2^*\ell\bar\nu) \over 
  {\cal B}(B\to D_1\, \ell\bar\nu) } =1.55\;.
\end{equation}
 We are now at the position to compare the theoretical predictions with the
 available experimental data. Recently the ALEPH and CLEO collaborations
 reported the measurement of the branching ratios for semileptonic $B$
 decay to excited $D$ mesons. For $B\to D_1\ell\bar\nu$ the average value
 of their results is
\begin{equation}\label{data}
  {\cal B}(B\to D_1\,\ell\bar\nu) = (6.5\pm 2.0) \times 10^{-3} \,.  
\end{equation} 
 The $B\to D_2^*\ell\bar\nu$ branching ratio has not yet been determined.
 CLEO set the limit ${\cal B}(B\to~D_2^*\,\ell\,\bar\nu)<0.8\%$ \cite{CLEO1}, 
while ALEPH found ${\cal B}(B\to D_2^*\,\ell\,\bar\nu)<0.2\%$ \cite{ALEPH1}.

Our prediction for $R$ is larger than available experimental data though we agree with the upper CLEO limit for 
$D^*_2$. This leads one to consider $O(1/m_{c,b})$ corrections.  One may 
consider the  mixing of
the $D_1$ with the $j_\ell=1^+/2$ meson, but it would worsen
our prediction since, due to the very small decay amplitudes into the
$j_\ell=1^+/2$ states, it would lessen our prediction for 
${\cal B}(B\to D_1\ell\bar\nu)$.
We are thus lead to consider direct $O(1/m_{c,b})$ corrections in the decay
amplitudes. A detailed discussion of $O(1/m_{c,b})$  corrections can be found
in Refs. \cite{Leib}, which  indeed enhance the rate to $B\to D_1\ell\bar\nu$
and lead to the expection that its branching ratio is greater than that for 
$B\to D_2^*\ell\bar\nu$. 

%%%%%%%%%%%%%%%%%%%%%%%%%%%%%%%%%%%%%%%%%%%%%%%%%%%%%%%%%%%%%%%%%%%%%%%%%%%
\section{Conclusion}
In this work we have presented the investigation for semileptonic $B$ decays into excited charmed
mesons. Within the framework of HQET we have evaluated the universal
Isgur-Wise functions $\tau(y)$ and $\zeta(y)$ by using QCD sum rules in
the leading order of $\alpha_s$. The $\tau(y)$ and $\zeta(y)$ functions
can be well fitted by linear approximations $\tau(y)(\zeta(y))=%
\tau(1)(\zeta(1))(1-\rho_{\tau(\zeta)}^2(y-1))$. The values of $\tau$
and $\zeta$ at zero recoil have been given. From a comparison with
the results of other approaches we found a broad agreement with some of
the quark model results for $\tau(1)$, whereas a smaller value for $\zeta(1)$.

We have computed, for the decays $B\to(D_1,D_2^*)\ell\bar\nu$ and
$B\to(D'_0,D'_1)\ell\bar\nu$, the differential decay widths and the
branching ratios. Our predictions for $D_1$ are  smaller than
experimental data, even take into account the wide spreading and
large uncertainty of experimental results. Our predictions for $D^*_2$
are below the experimental upper bounds of CLEO's results.  
We also predict the relation ${\cal B}(B\to D_2^*\ell\bar\nu)/%
  {\cal B}(B\to D_1\ell\bar\nu)$, which might be larger than available
  experimental limit. The discrepancy may be attributed to a large
  $1/m_c$ correction which enhances the transition to $D_1$. We predict tiny
branching ratios for $D'_0$ and $D'_1$, but there is no direct experimental check yet.

After finishing this paper we have learned of a paper by Colangelo $et$
$al.$ \cite{cnew},
in which the semileptonic decays $B\to (D'_0,D'_1)$ are studied by using a similar
approach. In particular, the radiative corrections are included in the
calculations of Isgur-Wise function $\xi$ in their work.
%%%%%%%%%%%%%%%%%%%%%%%%%%%%%%%%%%%%%%%%%%%%%%%%%%%%%%%%%%%%%%%%%%%%%%%%%%%%%%%%

\acknowledgments One of the Authors (M.Q.H) would like to acknowledge the
Abdus Salam International Centre for Theoretical Physics,
High Energy Group for hospitality. This
work was supported in part by
the National Natural Science Foundation of China.
\vspace{0.5cm}
%\begin{center}{\large{\bf Appendix }}\end{center}
\appendix
\section{}
We list here the  lorentz structures used in the paper.
\begin{eqnarray}
 {\cal L}^{\mu\nu}_{V}&=&\frac{1}{\sqrt{6}}\left[(y^2-1)g^{\mu\nu}+3v^\mu v^\nu+
 (1-2y)v'^\mu v'^\nu-(3yv^\mu v'^\nu+(y-2)v'^\mu v^\nu\right]\;,\\
 {\cal L}^{\mu\nu}_{A}&=&i\frac{1}{\sqrt{6}} (1+y)\epsilon ^{\mu\nu\alpha\beta}v_\alpha
 v'_\beta\;,\\
 {\cal L}^{\mu\alpha\beta}_{2V}&=&-\frac{i}{2}\left(\epsilon^{\mu\alpha\sigma\rho}
 (v_\beta-v'_\beta y)+\epsilon^{\mu\beta\sigma\rho}
 (v_\alpha-v'_\alpha y)\right)v_\sigma v'_\rho\;,\\
 {\cal L}^{\mu\alpha\beta}_{A}&=&\frac{1}{3}(1+y)g^{\alpha\beta}(v^\mu-v'^\mu)
 -\frac{1}{2}(1+y)g^{\alpha\mu}(v^\beta-v'^\beta y)
 -\frac{1}{2}(1+y)g^{\beta\mu}(v^\alpha-v'^\alpha y)\nonumber\\
    &&\quad\mbox{}
 +\frac{1}{2}(1-y)v'^\alpha v^\beta v'^\mu 
 +\frac{1}{2}(1-y)v^\alpha v'^\beta v'^\mu -\frac{1}{3}(1+y)
 v'^\alpha v'^\beta v^\mu\nonumber\\
    &&\quad\mbox{}
 -\frac{2}{3} (1+y)v'^\alpha v'^\beta v'^\mu
 +(v'^\alpha v'^\beta +v^\alpha v^\beta)v'^\mu\;,\\
 {\cal D}^\mu_A&=&v^\mu-v'^\mu\;,\\
 {\cal D}^{\mu\nu}_V&=&(y-1)g^{\mu\nu}+v'^\mu v^\nu-v'^\mu v'^\nu\;,\\
 {\cal D}^{\mu\nu}_A&=&-i\epsilon^{\mu\nu\alpha\beta}v_\alpha v'_\beta\;.
 \label{lorentz1}
\end{eqnarray}
  
%\newpage

\newpage
{\bf Figure Captions}
\vspace{2ex}
\begin{center}
\begin{minipage}{120mm}
{\sf Fig. 1.} \small{Feynman diagrams contributing to the sum rules for the 
                    Isgur-Wise form factors.  }
\end{minipage}\end{center}

\begin{center}
\begin{minipage}{120mm}
%\begin{minipage}{120mm}
{\sf Fig. 2.} \small{ Prediction for the  Isgur-Wise form factors. The upper 
band  corresponds to the $\tau$, the lower one to the $\xi$.}
\end{minipage}\end{center}
%\end{document}
\vspace{0.8cm}
%%%%%%%%%%%%%%%%%%%%%%%%%%%%%%%%table I here%%%%%%%%%%%%%%%%%%%
\begin{table}[htbp]
\caption{  Parameters of the form factors $\tau $, $\zeta$ }
\label{tab:comp}\vspace{2mm}
\centering
\begin{minipage}{11 truecm}
\begin{tabular}{ccccl}%\hline
%\hline%\hline
$\tau(1)$& $\rho^2_{\tau}$& $\zeta(1)$&
$\rho^2_{\zeta}$& Ref. \\
\hline
$0.74 $& $0.9 $& $0.26 $& $0.5 $& This work \\
$0.97$&$2.3$&$0.2$&$1.1$&\cite{DDG} \\
$0.71$& $1.5$& $0.8$& $1.0$& \cite{Leib} \\
$0.76$&$0.97$&$0.42$&$0.97$&\cite{dai2} \\
$0.49$& $0.9$&$0.5$& $0.4$& \cite{c-sum}\\
$0.54$& $2.8$& $0.62$& $2.8$& \cite{iw2}\\
$1.14$& $1.9$& $0.82$& $1.4$& \cite{wambach}\\
$0.89$& $1.45$& $0.12$& $0.73$& \cite{oliver},\cite{cccn}\\
$0.94$& $1.50$& $0.45$& $0.83$& \cite{oliver},\cite{godfrey}\\
% \hline%\end{minipage}
\end{tabular}\end{minipage}
\end{table}
%%%%%%%%%%%%%%%%%table II%%%%%%%%%%%%%%
 \vspace{3cm} 
\begin{table}[htbp]
\caption{Branching ratios for semileptonic  $B$ decays
 to P-wave charmed mesons  }
\label{tab:branch}\vspace{2mm}
\centering
\begin{minipage}{11 truecm}
\begin{tabular}{ccl}% 
Decay mode&Branching ratio &  
Exp. result \\
\hline
${B}\to D_{1}\ell\bar\nu$& $0.34$  & $0.56 \pm 0.13\pm 0.08\pm 0.04$
\cite{CLEO1}\\
${B}\to D^{*}_{2}\ell\bar\nu$&$0.52$ &$<0.8(90\%$ C.L.)\cite{CLEO1}\\
${B}\to D'_0\ell\bar\nu$&$0.019$  & -- \\
${B}\to D^{\prime}_{1}\ell\bar\nu$&$0.025$ &--  \\
\end{tabular}\end{minipage}
\end{table} 
%%%%%%%%%%%%%%%%Fig. 1%%%%%%%%%%%%%%%%%%55 
\begin{figure}[htbp]   % produce figure here 
\begin{center}
\setlength{\unitlength}{1truecm} 
\begin{picture}(6.8,6.8)%(<right,>top) 
\put(-8.0,-21)
{\includegraphics{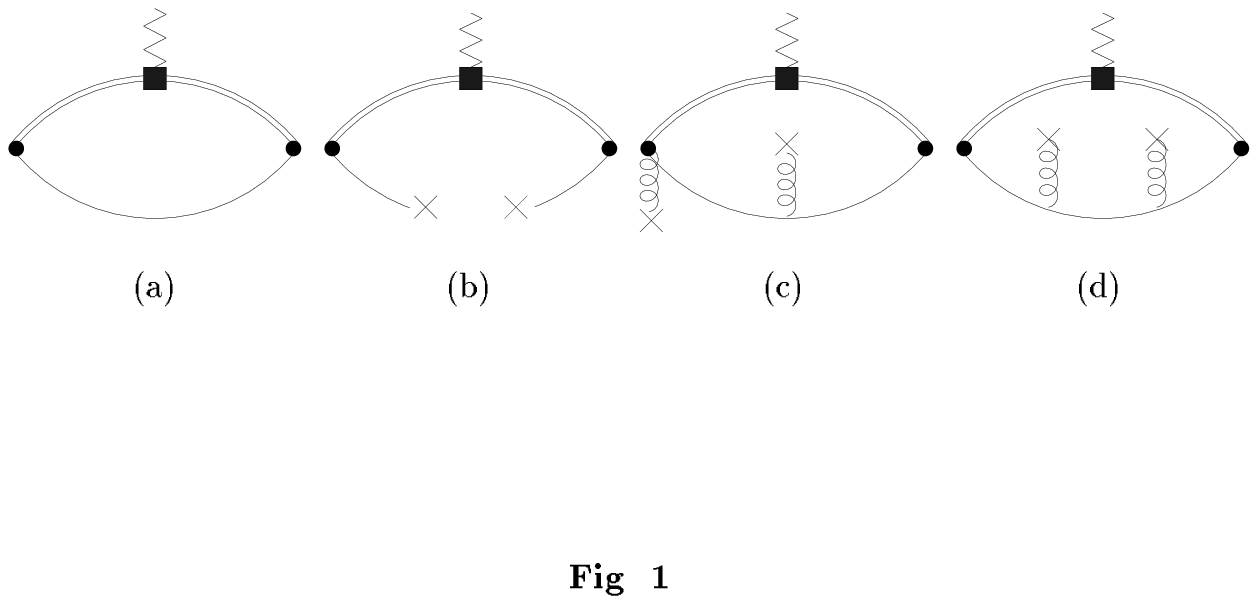}} 
\end{picture} 
\end{center} 
\vskip 2.0cm 
%\fcaption{xx}
\protect\label{Fig.1}
\end{figure}
%%%%%%%%%%%%%%%%Fig. 2%%%%%%%%%%%%%%%%%%%
\begin{figure}[htbp]   % produce figure here 
\begin{center}
\setlength{\unitlength}{1truecm} 
\begin{picture}(6.8,6.8)%(<right,>top) 
\put(-8.0,-14)
{\includegraphics{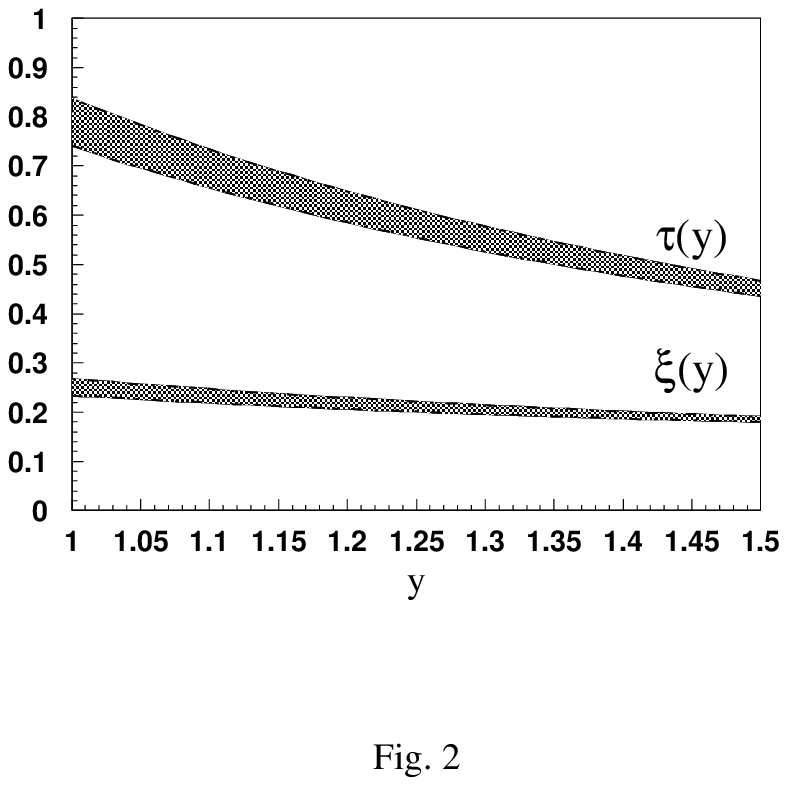}} 
\end{picture} 
\end{center} 
\vskip 2.0cm 
%\fcaption{xx}
\protect\label{Fig.2}
\end{figure}
 
\end{document}